\documentclass[a4paper,rmp,twocolumn,showkeys,superscriptaddress]{revtex4}

\usepackage{psfrag}        
\usepackage{amsmath}               
\usepackage[latin1]{inputenc}
\usepackage{graphicx} 
\usepackage{bm}
\usepackage{epstopdf}
\usepackage{amssymb}
\begin{document}
\def\k{\langle k\rangle}
\def\pa{\pi_a}
\def\pb{\pi_b}

\title{Is nestedness in mutualistic networks an evolutionary spandrel?}

\providecommand{\ICREA}{ICREA-Complex Systems  Lab, Universitat Pompeu Fabra,   Dr  Aiguader   88,   08003   Barcelona,   Spain}
   \providecommand{\ECLT}{European Centre for Living Technology, S. Marco 2940, 30124, Venice, Italy}
\providecommand{\WIEN}{Section for Science of Complex Systems, Med Univ. of Vienna, A-1090 Vienna, Austria}  
\providecommand{\IBE}{Institute of Evolutionary Biology, Psg Barceloneta, UPF-CSIC, Barcelona}    
\providecommand{\SFI}{Santa Fe  Institute, 1399 Hyde  Park Road, Santa Fe  NM  87501, USA}  
 \providecommand{\MICROSOFT}{Microsoft Research, Cambridge CB1 2FB, UK}
  \providecommand{\MOULIS}{Ecological Networks and Global Change Group, Experimental Ecology Station, Centre National de la Recherche Scientifique, 09200 Moulis, France}
\author{Sergi Valverde\footnote{corresponding   author}} \affiliation{\ICREA} \affiliation{\IBE} \affiliation{\ECLT} 
\author{Jose Montoya} \affiliation{\MOULIS}
\author{Lucas Joppa} \affiliation{\MICROSOFT}
 \author{Ricard  Sol\'e\footnote{corresponding   author}}\affiliation{\ICREA} \affiliation{\SFI} \affiliation{\IBE}


\begin{abstract}
Mutualistic networks have been shown to involve complex patterns of interactions 
among animal and plant species. The architecture of these webs seems to pervade 
some of their robust and fragile behaviour. Recent work indicates that there is a 
strong correlation between the patterning of animal-plant interactions and their 
phylogenetic organisation. Here we show that such pattern and other reported 
regularities from mutualistic webs can be properly explained by means of a very 
simple model of speciation and divergence. This model also predicts a co-extinction 
dynamics under species loss consistent with the presence of an evolutionary signal. 
The agreement between observed and model networks suggests that some 
patterns displayed by real mutualistic webs might actually represent evolutionary spandrels.
\end{abstract}
\maketitle

\section{Introduction} 

Ecological networks are well known to exhibit a number of structural features associated with their interaction patterns (1,2,3). 
Those include, in particular: 
(a) small world structure (4) where two given species are separated by a small 
number of links from any other species in the web (5,6,7), (b) 
heterogeneous distributions of connections (1,8) 
where the number of links between a given species and other species in the web can vary widely; 
(c) modular organisation (7,9) implying that subsets of species exhibit 
more connections among them than with the rest of the network and (d) nestedness (10), where specialists interact with a subset of the whole set of species that generalists interact with. 

The presence of some of these traits has important implications for the persistence and reliability of diverse ecosystems. As an example, it was shown earlier that the architecture of ecological food webs is consistent with the "robust-but-fragile" metaphor of complex networks: these webs are robust against the removal (extinction) of a random species but the removal of certain species can lead to a cascade of extinctions due to the existing chains of species dependencies (1, 11). 

Mutualistic networks describe species interactions across two adjacent trophic levels (of consumers and their resources), such as flowers and the insects that feed on and pollinate them. The bipartite graphs that these interactions form are considered the building blocks of biodiversity (12) and they are often significantly nested (13). Yet while the robustness of ecological networks might in large part be due to the presence of such structure, what is much less clear is from where or what that structure arises. 

Following an adaptationist view of naturally evolved systems, it has been argued that the presence of these properties, nestedness in particular, is a consequence of some underlying selection 
process that reduces competition relative to the benefits of facilitation and hence increases biodiversity and food web persistence or feasibility (3, 14, 15), although this has been challenged (16). 
The main arguments provided to support this view are grounded in the use of multispecies dynamical systems, based on generalised Lotka-Volterra equations with different functional responses. 
Several recent papers have questioned the conclusion that nestedness  
has resulted from selection pressures (17) favouring higher biodiversity. 
Instead, it has been suggested that nestedness is likely to be a consequence (instead of a causative property) of biodiversity, 
in particular of the heterogeneous distributions of connections. 

Other studies seem to support this view, where a structural pattern is incorrectly pointed to as a causal agent for a given 
functional trait and the biological details of the system under consideration. 
In this context, previous work concerning the evolution of complex biological 
and artificial networks suggest that many architectural patterns displayed 
by these graphs are an inevitable byproduct of the way they are constructed (18). 
This is in fact the consequence of processes involving network growth through 
duplication and rewiring (19, 20, 21, 22, 23). Specifically, evolution often proceeds by tinkering from available components (24, 25) and a network 
(including the proteome, metabolic networks and even technological graphs) resulting from a process 
of copy and further modification is likely to display complex features. Simple models involving no functionality or population dynamics can develop small world or scale-free webs, which can be modular (26) despite the apparently well established idea that modularity is an evolved, functionally relevant trait. 
If this were the case for mutualistic webs, their invariant features (27) would be  a consequence of universal properties of the graphs and their growth rules, closer to the idea of {\em universality} (28, 29) . 
When this occurs, very simple, toy models are capable of accounting for the global features exhibited by the system.

The key lesson of the studies mentioned above is that, 
when dealing with complex biological networks, it is important to consider the generative rules responsible for their growth and change. 
Some of these very ubiquitous patterns might be a byproduct of these rules, although they might be relevant 
or even functionally important afterwards. The emergent patterns can thus be {\em evolutionary spandrels}, i.e., 
phenotypic characteristic that evolved as a side effect of a true adaptation (30,31). 
Despite some criticisms related to the appropriateness of the architectural analogy (32) the key concept of a non-adaptive structural patterns stands. An example of spandrel is provided by the distribution of network motifs in cellular networks (33) where it has been shown that the conserved, uneven distribution of some small subgraphs can be explained by means of non-functional models. 
We can define evolutionary spandrels as structures that: (i) are the byproduct of building rules; (ii) have intrinsic, well-defined, non-random features; and (iii) their structure reveals some of the underlying rules of
construction (33).
 
\begin{figure}
\begin{center}
\includegraphics[scale=0.55]{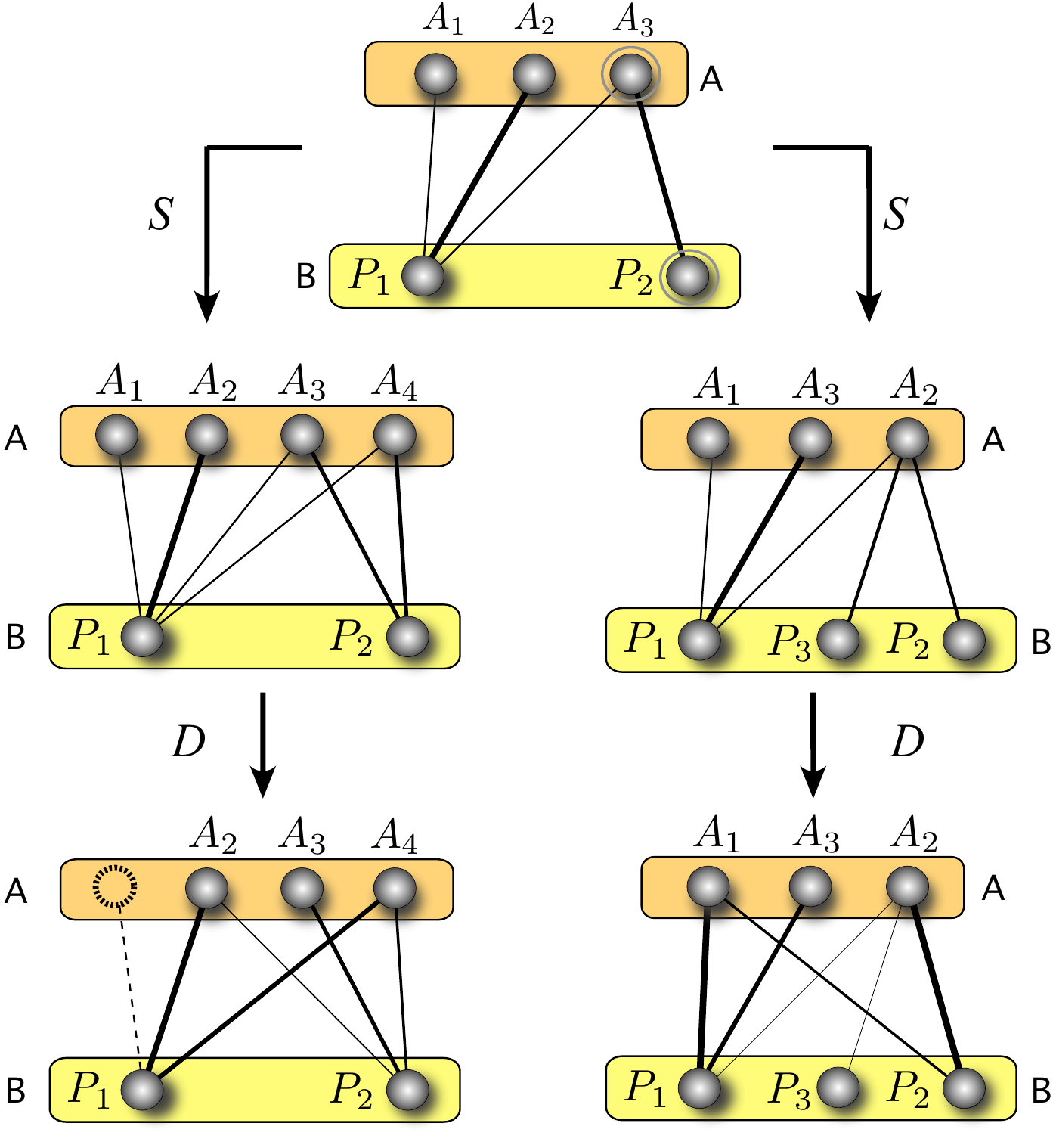}
\caption{Duplication-divergence rules: (a) The graph is composed of two 
layers involving animals (upper) and plants (lower). Speciation can affect either the $A$ or the $P$ sets (left and right sequences, respectively). Each time a new species is added (b) the daughter inherits all its interactions. Afterwards, they experience a divergence (c) affecting the weights and links.} 
\label{fig1}
\end{center}
\end{figure}

Here we aim to show that nestedness, as well as other key statistical features of mutualistic webs, can be recovered as a byproduct of the generative rules associated with the creation of diversity through speciation-divergence dynamics with no consideration of the underlying population dynamics. 
This approach ignores the ecological time scale (and thus all factors associated to standard stability criteria) by considering instead a scenario 
where speciation and diversification that takes place over very long (evolutionary) time scales. 
Such kind of model has been used to model macroevolutionary dynamics using both adaptive dynamics on fitness landscapes as well as models of network growth and extinction (34,35,36,37,38,39,40). 
These class of models has been able to give insight into the 
large-scale evolution of ecological networks (41, 42 and references therein). 
A crucial point of using these models is that we can explore the outcome of evolutionary rules that drive the structural patterns of connectivity beyond the ecological time scale.

\section{Speciation-diversification model} 

The approach taken here makes some strong assumptions. One is that species are either present or absent, 
with no role to be played by population size or other species-specific traits. Secondly, interactions 
are introduced as weighted links. The values of these links will evolve in time following very simple 
rules. The large-scale dynamics of our system is obtained by a combination of two processes that obviously 
occur over evolutionary time scales: new species are generated from old ones through speciation and 
coevolution and external (either environmental or stochastic) factors modify the presence and strength of the interactions. 

\begin{figure*}[htpq]
\begin{center}
\includegraphics[scale=0.9]{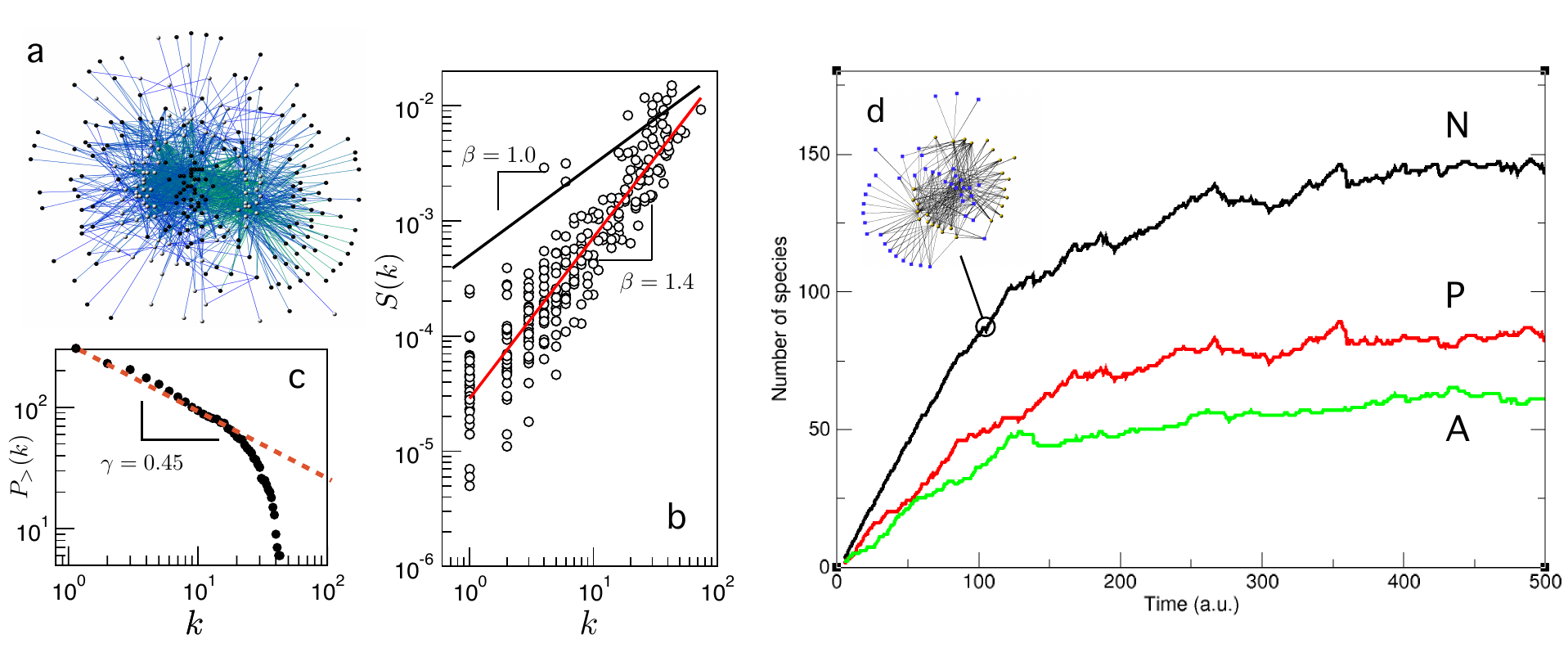}
\caption{(a) an example of the structured, heterogeneous network resulting 
from the growth dynamics with speciation and divergence (see text). Here dark and light nodes indicate "plants" and "animals". In this example, $N=267$ species where present at the end of the simulation, after $T=500$ speciation-divergence steps. The strength of the links is shown by a color scale. Greener and bluer links indicate stronger and weaker connections, respectively.  In (c) the undirected degree distribution of the graph shown in (a) is displayed in a log-log plot. We can see that a the system follows a broad distribution. 
The plot (d) shows the strength-degree correlation, which gives a power law $S(k) \sim k^{\eta}$ with $\eta=1.4 \pm 0.05$.  The linear expectation is also shown (black line) for comparison. Here we have used $\beta=10^{-5}, p=0.1$ and $\theta=10^{-6}$. An example of the time series generated by the model is shown in (d) with a snapshot of the whole graph at some intermediate time (here $t=100$).} 
\label{fig2}
\end{center}
\end{figure*}

Our model assumes a bipartite graph $G$ involving two subsets $A(t)$ and $P(t)$ that correspond 
to the animals and plants, respectively, at a given evolutionary time step t. These species are linked (figure 1) 
provided that a mutualistic relationship exists. Here we  consider the effect of animals on plants, weighted through a matrix 
$W_{ij}=W(A_i \rightarrow P_j)$ that indicates the strength of the interaction between both partners. This 
can be interpreted in terms of the number of dependencies existing between the given pair (in one direction). 
The evolutionary dynamics are defined by a simple set of rules:
\vspace{0.2 cm}
\\
\noindent
{\bf Speciation}: we choose a given species $A_i$ or $P_j$  and create a speciation event. 
The new species inherits exactly the same list of links from its parent species. 
If $A_k$ indicates the newly created species, then we have $W_{kj}=W_{ij}$.
\vspace{0.2 cm}
\\
{\bf Divergence}: we redistribute the weights between parent and daugther species. 
A random number $0<\mu<1$ is generated and each pair of links $\{ W_{kj},W_{ij} \}$ is updated to a new pair 
$\{\mu W_{kj}, (1-\mu)W_{ij} \}$.
Additionally, for each link, we introduce, with a given probability $p$,  a weight change, i. e., we have a new value
\begin{equation}
W_{ij} \rightarrow W_{ij} \pm \xi
\end{equation}
with $\xi$ being a small random number  between zero and $\beta$. Here the parameter $\beta$ will weight 
how fast evolutionary changes occur at the level of single ecological links. If $W_{ij}$  
falls below a threshold $\theta$, it is removed. 
Finally, a maximum input weight is allowed for all plants. Specifically, if the sum 
\begin{equation}
S_j = \sum_i W(A_i \rightarrow P_j) 
\end{equation}
over all animals acting on the plant $P_j$  is larger than one, the change is not accepted. A symmetric rule 
is used to constraint the links in the $P \rightarrow A$ direction. 

Because of the type of dynamics defined by the previous rules, species become extinct when 
no mutual support is present (i. e. if $\sum_j W_{ji}=0$). This is the simplest way of defining the 
mutual cooperation among species.

\begin{figure*}[htpq]
\begin{center}
\includegraphics[scale=0.8]{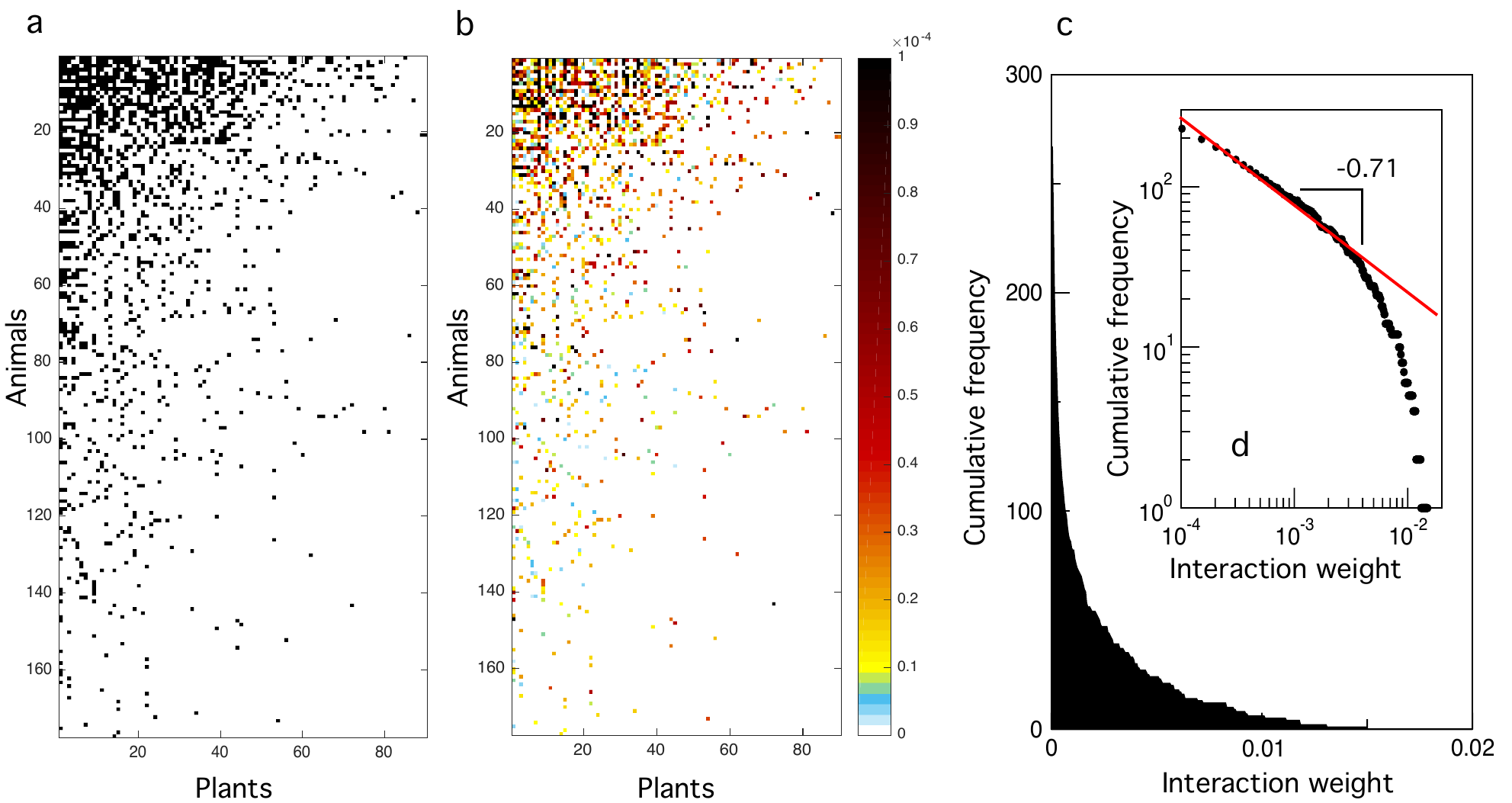}
\caption{The matrix of interactions between the two layers of our bipartite graph model 
of mutualistic interactions. In (a) we display the adjacency matrix, where black points indicate the 
presence of a connection. Arrows and columns have been Figure (b) show the weights of these links. 
In (c) the distribution of link weights is displayed, again exhibiting 
a trncated power law, i. e. $P(s) \sim s^{-\gamma_s} \exp (-{s /s_c})$. } 
\label{fig2}
\end{center}
\end{figure*}

The model successfully generates networks with all the reported statistical patterns displayed by empirical webs. 
In figure 2a-b we show an example of a graph obtained from our rules, starting from an initial 
condition with two species at each level, connected to each other with a small 
weight $W_{kl} = 10^{-3}$. The webs are heterogeneous, showing broad scale distributions of 
connections (figure 2c). Specifically, if $k^a_i$ and $k^p_j$ indicate the number of links (or degree) of 
$A_i$ and $P_j$, respectively, the frequency of species having a given number of links $k$ 
will be indicated by $P(k)$. It has been found that for mutualistic webs the appropriate form of 
these distributions is (27): $P(k) \sim k^{-\gamma} \exp (-{k /k_c})$ 
where $\gamma$ is the exponent that indicates how rapidly the distribution falls at small $k$ 
and $k_c$ a cut-off that effectively limits the spread of the distribution (43). 
The larger $k_c$ the more flat is the distribution and the more common the presence of highly-connected species. 
In figure 2c we represent the cumulative degree distribution $P_>(k)$ defined as:
 \begin{equation}
P_>(k) = \int_k^{\infty} P(k') dk'
 \end{equation}
This distribution allows to smooth out the fluctuations and to define better estimates of the two characteristic parameters $\gamma$ and $k_c$. Our model correctly predicts the broad distribution scenario, thus indicating that heterogeneity is a consequence of amplification-divergence phenomena, consistently with previous studies (19) but in this case applied to a bipartite system. 

Another property of the network that can be easily determined is the aggregated {\em strength} of the nodes, 
defined as the sum of all the dependencies in both directions i. e. 
\begin{equation}
S_i  = \sum_{j=1}^{N_i}  ({W_{ij}+W_{ji}})
\label{strength}
\end{equation}
Where $N_i$ indicates the number of interactions with other species. 
The use of a weighted network provides relevant information 
about its local and global organisation (44,45,46) and was early identified as an essential feature of mutualistic interactions (47). 
In general, a scaling law relates strength and degree, namely 
\begin{equation}
S(k) \sim k^{\eta}
\end{equation}
where the exponent $\eta$ will establish the nature of the correlation. For a randomly distributed 
set of weights, it can be shown that a linear relation ($\eta=1$) exists. However, if the importance of a given 
node in the network is lower than predicted by its degree, then we would observe $\eta<1$. 
In mutualistic webs, it has been shown that a superlinear behaviour is found, i. e. $\eta>1$, indicating that 
species with many connections tend to display stronger interactions than the average. This is the case of 
real webs and also occurs in our model, as displayed in figure 2d. 
In our simulated system, we have $\eta=1.40 \pm 0.05$ which is also close to measured data (45).

The previous patterns of network organisation reveal that some key statistical regularities found in mutualistic webs can be obtained from our minimalistic assumptions. 
These results are the first indication that the large-scale organisation of mutualistic webs could be a side effect of the amplification dynamics associated to the copy-and-divergence dynamics associated to the rules considered here. 
What about nestedness? Is this property also an emergent phenomenon resulting from evolutionary dynamics, decoupled from the underlying ecological dynamics?

\section{Nestedness for free}

In this section, we study the nestedness property associated our model. 
We show that our  model exhibits nestedness without any additional stability requirement. 
A quantitative (weighted) approach to nestedness (unlike binary measures) takes
 into account species abundances and interaction frequencies. 

A bipartite network $G=(A,P,E)$ has two disjoint sets of nodes $A$ and $P$  representing animals and plants, respectively. Let $N_A=|A|$ be the number of animal species and $N_P=|P|$ be the number of plant species, i.e., the total number of species in our system is $N = N_A + N_P$.
Now, assume that animals are indexed $1,2,..,N_A$ and plants are labeled $N_A+1, N_A+2, ... ,N_A+ N_P$. 
The matrix of mutualistic interactions $W=[ W_{ij} ]$ has a block off-diagonal form like this:
\begin{equation}
W=\begin{bmatrix} 0 & { W }_{ { N }_{ A }\times { N }_{P} } \\ \\
\\
{ W }_{ { N }_{P}\times { N }_{ A } } & 0 \end{bmatrix}
\end{equation}
  

where $0$ is the all-zero matrix that reflects the bipartite constraint, i.e., there are no interactions between any pair of alike species. 
The  so-called {\em bipartiteness} function $\nu (u)$ indicates $u$ node  type, e.g.,  where the vertex belongs to one type ($\nu(u)$ = 0) or the other type ($\nu(u)$ = 1). 
Using this notation, bipartite edges $(u,v) \in E$ must satisfy $\nu(u) + \nu(v)$ = 1.  

Edges in a nested network are organised in a way that specialists interact with subset of the species whom generalists interact with. 
This nested pattern can be detected in the specific arrangement of present and absent interactions in bipartite networks. 
Recently, this nestedness definition was extended to quantitative networks using spectral graph theory. 

It can be shown that the largest eigenvalue of bipartite networks determines nestedness (16). 
This is a robust measurement of nestedness because its invariance to the sorting of rows and columns in the matrix.
First, we define the  $N_A \times N_P$ incidence matrix $B=[B_{ij}]$: 
\begin{equation}
B_{ij} = {{1}\over{2}} ( W_{i,j+N_A} + W_{j+N_A, i} )  
\end{equation}
where $1 \le i \le N_A$ and $1 \le j \le N_P$. We can interpret
this bipartite matrix as the average interaction frequency
between a pair of (animal, plant) after discarding link direction. 
Because $B$ is a symmetric matrix, all its eigenvalues are real and 
distributed symmetrically around 0.  The  spectral radius $\rho(B)$ (or
dominant eigenvallue) is the largest eigenvalue associated to the matrix $B$ 
and it represents a natural measurement of nestedness: large values 
of $\rho(B)$ correspond to highly nested matrices.

Nestedness is a relative value that depends on the
 size (number of species $N$) and fill (density of interactions)
of the bipartite matrix $B$. In order to assess its relevance, 
we compare the observed value of nestedness in the model with an ensemble of 
random matrices with similar properties (48, 49). Here, we use
the null model proposed by (16), which keeps the structural 
features of the network while swapping the order of weighted links 
(so-called 'binary shuffle' in (49)).
We assess the significance of empirical nestedness with the Z-score: 

\begin{equation}
Z={{\rho(B) - \left < \rho \right >} \over {\sigma_\rho}} 
\end{equation}

where $\left < \rho \right >$ and $\sigma_\rho$ are the average value and the standard deviation of the network measure in a random ensemble, respectively. Here, we consider that mutualistic  networks are significantly nested whenever the corresponding $Z > 2$ (i.e., $p < 0.05$ using the Z-test). 

\begin{figure}[t]
\begin{center}
\includegraphics[scale=0.5]{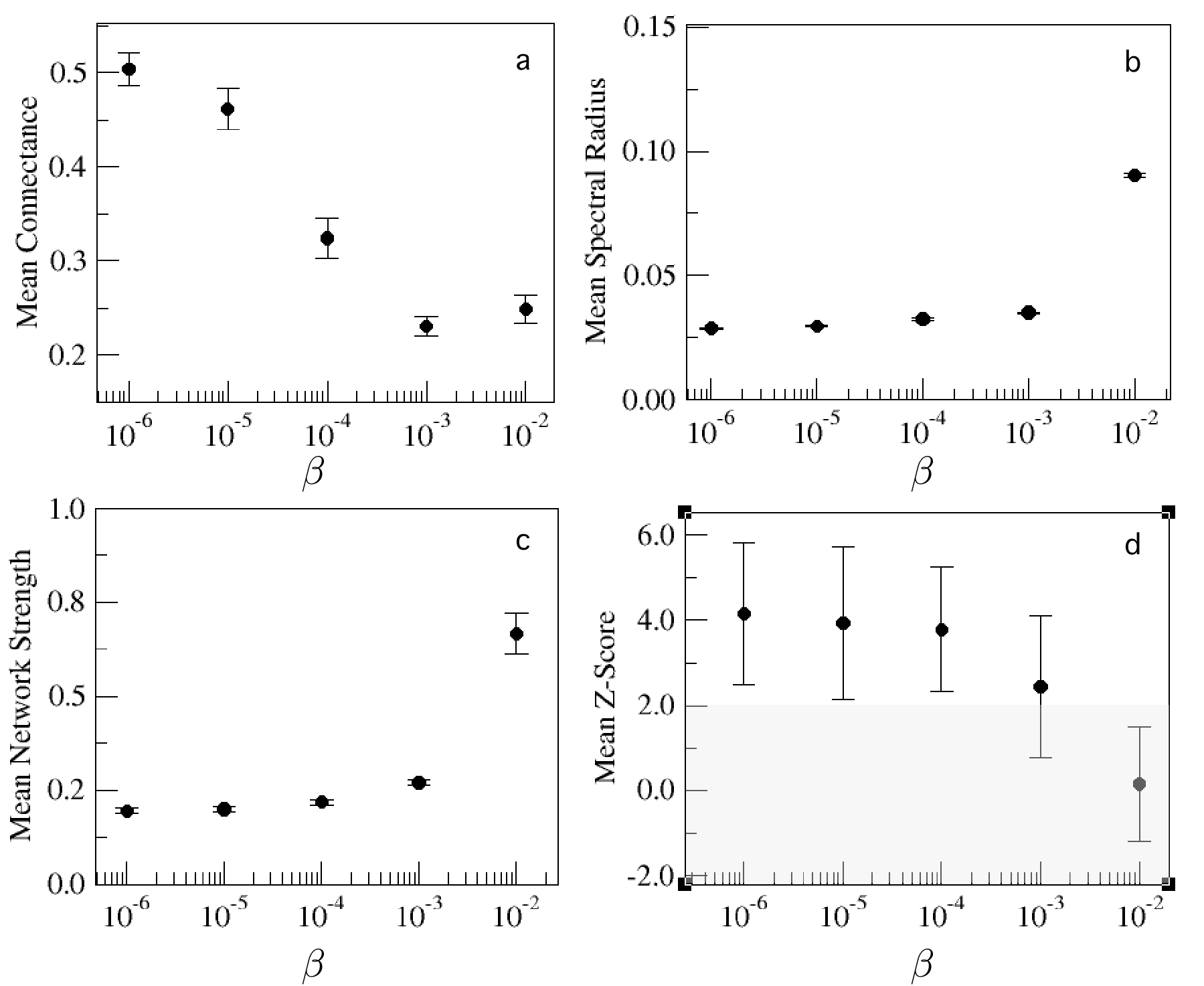}
\caption{The effect of the maximum link change $\beta$ on the structural patterns exhibited by mutualistic model networks: (a) mean connectance decays with faster evolutionary rates, (b) nestedness is stable for wide range of parameters and (c) it correlates with network strength, (d) statistical signficance of the nestedness drops for high evolutionary rate, i.e., $\beta > 0.01$.  Here we have used link probability $p \in [0,1]$ and tuned $\beta$ over several orders of magnitude. Overall, these plots comprise 600 random samples taken uniformly from the phase space $(p, \beta)$. 
} 
\end{center}
\end{figure}


What is the impact of parametric changes on the statistical properties of these webs? 
A systematic exploration reveals that the previous conclusions are robust.
The main  parameter associated to our model is $\beta$, which provides a measure of the 
allowed speed of network changes through evolutionary time. We have explored 
the impact of this parameter on the connectance, mean network strength and 
spectral radius and nested organisation of these webs (see figure 4).
Our model predicts highly significant nestedness for a 
wide range of parameters, i.e., when $\beta < 0.001$. On the other hand, 
the grey region in Figure 4d corresponds to  distributions of spectral
radius scores in an ensemble of random matrices. The results
described here support the hypothesis that and evolutionary-scale, minimal  
model generates highly nested bipartite networks in a wide range of
scenarios.

\section{Discussion}

In this paper we have introduced a very simple model of large scale evolution of 
mutualistic webs. It includes the most elemental requirements associated to the 
measurable web properties. It ignores all details except minimal components related to 
the creation of new species through speciation and the change (under constraints) of 
weights over evolutionary time. Despite its simplicity, the model is capable of 
consistently reproducing several well known structural patterns of organisation, suggesting that 
the generative rules responsible for network growth largely determine the presence of 
universal traits in empirical systems. The model incorporates a restriction to the total amount of inputs received 
and the persistence of species is guaranteed provided that a link exists between this 
species and at least one mutualistic partner. 

Generative models of network structure are seldom rare mutualistic 
networks in comparison with other ecological networks, e.g. antagonistic ones.
The model presented here is not the only one trying to explain the emergence 
of complex mutualistic networks. In (50) the authors showed that nestedness and 
heterogeneous degree distributions emerge from an optimization principle that maximizes species abundances. Their model, however,
works over ecological time-scales, while the question we asked here is to what extent simple evolutionary models can account for 
observed structural patterns. In this respect, Nuismer et al (2013) developed a quantitative genetic model that allows inference of network structure over evolutionary time.Their model was far more complex that the one presented here, 
but they only found that emerging networks were more nested than their random 
counterparts under very restrictive conditions: species interactions 
should be mediated by phenotype differences and coevolutionary selection should
be weak. In the remaining cases, resulting networks were either not nested or anti-nested.

Duplication-rewiring models are known to indirectly incorporate a preferential attachment rule. 
This rule is known to generate heterogeneous, sometimes long-tailed degree distributions (51). 
Once heterogeneous distributions arise, other features can come "for free": nestedness in particular has been shown to be largely a consequence of broad connectivities (52, 53). 

Our model incorporates evolutionary rules of speciation and drift that naturally provide a mechanism to explain the properties found in mutualistic webs. 
Since our model does not include the population size associated to each species nor the nonlinear dynamics of ecological interactions, our results suggest that there is no need to assume that the ecological scale plays a major role in shaping mutualistic webs. 
Instead, the universal constraints associated to the evolutionary unfolding of these webs would lead to the observed invariant properties. 

\section*{Acknowledgments}

The authors thanks Stuart Pimm and the members of the Complex Systems Lab for useful discussions. 
This work was supported by Bot\'in Foundation by Banco Santander through its
Santander Universities Global Division. We also thank the Centre of Living technology and the Santa Fe Institute, where most of this work was done. This work was supported by the Spanish Ministry of Economy and Competitiveness, Grant FIS2016-77447-R and FEDER (SV). JMM acknowledges support by the 
French Laboratory of Excellence project TULIP (ANR-10-LABX-41; ANR-11-IDEX-002-02).

\begin{enumerate}
\item  
Sol\'e, R.V.; Montoya, J.M. 2001. Complexity and fragility in ecological networks.  Proc. Roy. Soc. Lond. Ser B 268, 2039-2045.

\item 
Montoya, J.M., Pimm, S. and Sol\'e, R. 2006. Nature 442 (7100), 259-264.

\item 
Bastolla, U. et al. 2009. The architecture of mutualistic networks minimizes competition and increases biodiversity. 
Nature 458: 1018-1020.

\item 
Watts, D. J. and Strogatz, S. H. 1998. Collective dynamics of "small-world" networks. Nature 393: 440-442.

\item 
Montoya, J.M. and Sol\'e, R.V. 2002 Small world patterns in food webs. J. Theor. Biol. 214, 405-412.

\item 
Dunne, J.A.; Williams, R.J.; Martinez, N.D. 2002. Food-web structure and network theory: The role of 
connectance and size. Proc Natl Acad Sci USA, 99, 12917-12922.

\item 
Olesen, J. et al. 2006. The smallest of all worlds: pollination networks. Ecological networks and their fragility. 
J. Theor. Biol. 240,  270-276.

\item 
Bascompte, J. et al. 2006. Asymmetric coevolutionary networks facilitate biodiversity maintenance. Science 312: 431-433.

\item 
Fortuna, M.A., Stouffer, D.B., Olesen, J.M. et al. 2010.
Nestedness versus modularity in ecological networks: two sides of the same coin?
J. Anim. Ecol. 79, 811-817.

\item 
Bascompte, J., Jordano, P., Meli\'an, C.J. and Olesen, J.M. 2003. The nested assembly of plant-animal mutualistic networks. 
Proceedings of the National Academy of Sciences USA, 100, 9383-9387.

\item 
Memmott, J. et al. 2004. Tolerance of pollination networks to species extinctions. Proc. R. Soc. B 271, 2605-2611.

\item 
Joppa, L.N. and Williams, R. 2013. 
Modeling the building blocks of biodiversity. PLOS ONE 8, e56277.

\item 
Joppa, L.N., Montoya J.M., Sol\'e, R., Sanderson, J., and Pimm, S.L. 2010. On nestednes in ecological networks. Ecol. Evol. Res. 12, 35-46.

\item 
Stouffer, D.B., and Bascompte, J. 2011.
Compartmentalization increases food-web persistence
Proceedings of the National Academy of Sciences (108), 9,  648-3652.

\item 
Saavedra, S., Rohr, R. P., Olesen, J. M., and Bascompte, J. 2016. 
Nested species interactions promote feasibility over stability during the assembly of a pollinator community. 
Ecology and evolution. 6, 997-1007.

\item 
Staniczenko, P. P. A.; Kopp, J. C.; Allesina, S. 2013. 
The ghost of nestedness in ecological networks. Nature Comm. 4: 1391.  

\item 
James, A., Pitchford, J. W. and Plank, M. J. 2012. 
Disentangling nestedness from models of ecological complexity. Nature 487, 227-230.

\item 
Lynch, M. 2007. The evolution of genetic networks by non-adaptive processes. Nature Rev. Genet. 8, 803-813.

\item 
Sol\'e, R., Ferrer Cancho, R., Montoya, J. and Valverde, S. 2003. Selection, 
Tinkering, and Emergence in Complex Networks. Complexity 8(1), 20-33.

\item 
Banzhaf, W. and Kuo, P.D. (2004) Network motifs in natural and
artificial transcriptional regulatory networks. J. Biol. Phys. Chem. 4, 85-92

\item 
Mazurie, A. et al. 2005. An evolutionary and functional assessment of regulatory network motifs. Genome Biol. 6, R35.

\item 
Zhang, L.V. et al. (2005) Motifs, themes and thematic maps of an integrated {\em Saccharomyces cerevisiae} 
interaction network. J. Biol. 4 doi: 10.1186 jbiol23

\item 
Rodriguez-Caso C., Medina M. A., Sol\'e R.V. 2005.
Topology, tinkering and evolution of the human transcription factor network.
FEBS J. 272 (24), 6423-6434.

\item 
Jacob, F. 1977. Evolution and Tinkering. Science 196, 1161-166.

\item 
Sol\'e, R.V.; Pastor-Satorras, R.; Smith, E.; Kepler, T.  2002. 
A model of large-scale proteome evolution. Adv. Complex Syst. 5, 43-54.

\item 
Sol\'e, R.V., and Valverde, S. 2008. Spontaneous emergence of modularity in cellular networks.
Journal of the Royal Society Interface 5 (18), 129-133.

\item 
Jordano, P. et al. 2003. Invariant properties in coevolutionary networks of 
plant-animal interactions. Ecol. Lett. 6: 69-81.

\item 
Ward M. 2001. {\em Universality}. McMillan, London. 

\item 
McComb, W. D. 2008. {\em Renormalization Methods: A Guide For Beginners}. Oxford U. Press. 

\item 
Gould, S.J. and Lewontin, R.C. (1979) The Spandrels of San Marco and the 
Panglossian Paradigm: a critique of the Adaptationist Programme. Proc. R. Soc. B 205, 581-598

\item 
Gould, S.J. (2002) {\em The Structure of Evolutionary Theory}, Harvard University Press. 

\item 
Dennett, D. C. 1995. {\em Darwin's dangerous idea}. Simon and Schuster, New York. 

\item 
Sol\'e R. and Valverde, S. 2006. Are network motifs the spandrels of cellular complexity?
Trends Ecol. Evol. 21(8), 419-422.

\item 
Kauffman, S.A. and Levin, S. 1987. 
Towards a general theory of adaptive walks on rugged landscapes. J Theor Biol 128, 11-45.

\item 
Kauffman, S.A. and Johnsen, J. 1991. Coevolution to the edge of chaos: coupled fitness landscapes, 
poised states and coevolutionary avalanches. J. Theor. Biol. 149, 467-505.

\item 
Sol\'e, R.V. and Manrubia, S. C. 1995. Extinctions and self-organised criticality in a model of 
large-scale evolution. Phys. Rev E 54, R42-R46.

\item 
Caldarelli, G., Higgs, P. G. and McKane, A. J. 1998. 
Modelling Coevolution in Multispecies Communities. J. Theor. Biol. 193, 345-358.

\item 
Christensen, K., Di Collobiano, S. A., Hall, M., and Jensen, H. J. 2002. 
Tangled nature: a model of evolutionary ecology. J. Theor. Biol. 216, 73-84.

\item 
Newman, M. E. J. and Palmer, R. 2003. {\em Modelling extinction.} Oxford U. Press. New York.

\item 
Dorogovtsev, S.N. and Mendes, J.F.F. 2002. Evolution of Random Networks. Adv. Phys.  51, 1079-1187.

\item 
Loreau, M. 2010. Linking biodiversity and ecosystems: towards a unifying ecological theory. 
Phil. Trans. R Soc. B 365, 49-60.

\item 
Sol\'e, R. V., and Bascompte, J. 2006. 
{\em Self-organization in complex ecosystems}. 
Princeton U. Press. 

\item 
Amaral, L.A.N.; Scala, A.; Barthelemy, M. and Stanley, H.E. 2000. 
Classes of behavior of small-world networks. Proc. Nat. Acad. Sci. USA, 97, 11149-11152.

\item 
Barrat, A., Barth\'elemy, M., Pastor-Satorras, R., and Vespignani, A. 2004.
The architecture of complex weighted networks
Proceedings of the National Academy 101(11), 3747-3752.

\item 
Bascompte, J. and Jordano, P. 2007. Plant-animal mutualistic
networks: the architecture of biodiversity. Annu. Rev. Ecol. Evol. Sydney. 101, 221-223.

\item 
Gilarranz L. J., Pastor, J. M. and Galeano, J. 2012. The architecture of weighted mutualistic networks. 
Oikos 121, 1154-1162.

\item 
Jordano, P. 1987. Patterns of mutualistic interactions in pollination and seed dispersal: 
connectance, dependence asymmetries and coevolution. Am. Nat. 129, 657-677.

\item 
Weitz, J. S.; Poisot, T.; Meyer, J. R.; Flores, C. O.; Valverde, S.; Sullivan, M. B.; Hochberg, M. E. 2013. Phage-bacteria infection networks. Trends in Microbiology 21(2), 82-91.

\item 
Beckett, S.; Boulton, C. A., Williams, H. T. P. 2014. FALCON: a software package for analysis of nestedness in bipartite networks. F1000Research 3, 185. 

\item 
Suweis, S., Simini, F., Banavar, J. R. and Maritan, A. 2013. 
Emergence of structural and dynamical properties of ecological mutualistic networks. Nature, 500, 449-452.

\item 
Vazquez, A. 2003. Growing network with local rules: Preferential attachment, clustering hierarchy,
and degree correlations. Phys. Rev E67, 056104.

\item 
Jonhson S, Dominguez-Garcia V. and Munoz, M. A. 2013. 
Factors Determining Nestedness in Complex Networks. PLOS ONE 8, e70452.

\item 
Feng, W. and Takemoto, K. 2014. Heterogeneity in ecological mutualistic networks 
dominantly determines community stability. Sci. Rep. 4: 5219.

\end{enumerate}

\end{document}